# Is there a resting frame in the universe? A proposed experimental test based on a precise measurement of particle mass

Donald C. Chang

Hong Kong University of Science and Technology, Clear Water Bay, Hong Kong
Email: bochang@ust.hk

**Abstract**

*According to the Special Theory of Relativity, there should be no resting frame in our universe. Such an assumption, however, could be in conflict with the Standard Model of cosmology today, which regards the vacuum not as an empty space. Thus, there is a strong need to experimentally test whether there is a resting frame in our universe or not. We propose that this can be done by precisely measuring the masses of two charged particles moving in opposite directions. If all inertial frames are equivalent, there should be no detectable mass difference between these two particles. If there is a resting frame in the universe, one will observe a mass difference that is dependent on the orientation of the laboratory frame. The detailed experimental setup is discussed in this paper.*

**Keywords**: vacuum; relativity; resting frame; particle mass; mass spectrometry

PACS: 03.30.+p; 07.75.+h

## 1 Introduction

In the famous experiment conducted by Michelson and Morley in late 19th century, it was found that the propagation of light is independent of the movement of the laboratory system [1]. This finding was interpreted by Einstein as an indication that the physical laws governing the propagation of light are equivalent in all inertial frames. In a paper published in 1905, Einstein raised this understanding to the status of a postulate: "…*the same laws of electrodynamics and optics will be valid for all frames of reference*…" [2]. This postulate was known as the 1$^{st}$ postulate of the Special Theory of Relativity (STR).

One may notice that this postulate of relativity was originally applied only to "*electrodynamics and optics*". But later, this postulate was generalized to all physical laws [3]. This thus raises a serious question: Can this generalization be justified? The results of Michelson-Morley experiment only demonstrated that the propagation of light obeys the principle of relativity. What about particles with rest mass? Can one demonstrate that the physical behavior of massive particles also obey the 1$^{st}$ postulate of relativity?

This postulate of relativity implies that there is no resting frame in our universe; otherwise one will be able to determine which inertial frame is stationary and which frame is moving. This means that the vacuum in our universe must be an empty space and thus cannot serve as a reference system. Such a requirement, however, will be in conflict with the modern view of the vacuum. In the Standard Model of cosmology today, the vacuum is far from being empty. Many cosmology theories assume that the energy of our universe comes from the quantum fluctuation in the vacuum [4], which clearly cannot be considered as an empty space. Also, according to the most recent CMB (cosmic microwave background) studies, our universe is filled with visible matter, dark matter and dark energy [5, 6]. Could they form a resting frame?

An empty vacuum is also not consistent with the current theories of particle physics. For example, in quantum electrodynamics, every oscillation mode is supposed to have a zero-point energy [7]. Such energy is treated as a part of the vacuum system. In fact, in the quantum field theory, the vacuum is always regarded as the ground state. The physical fields are just excitations above the vacuum [8]. Should this vacuum form a resting frame?

## 2 Principle of the experimental test for detecting the motion of an inertial frame

In this work, we would like to propose an experimental test to determine whether there is a resting frame in our universe or not. The basic idea is to measure the speed-dependency of the mass of a charged particle in different directions. From earlier experimental studies, we know the mass of a particle is not constant; it is dependent on its traveling speed [9], such that,

$$M = \frac{m_0}{\sqrt{1 - v^2/c^2}},  \quad (1)$$

where $M$ is the moving mass and $m_0$ is the rest mass. When the speed of the particle increases, the moving mass of the particle will also increase. In this experiment, we will measure the masses of two identical particles travelling in opposite directions with the same speed ($u'$). If all inertial frames are equivalent with respect to all laws of physics, the measured masses for the two particles should be the same. But if there is a resting frame in our universe, the situation will be different. Suppose the laboratory is traveling at a speed $v_0$ relative to the resting frame along the *x*-axis (see **Fig. 1**), then the speeds for the two particles are not equal according to the resting frame. The particle moving to the right should have a higher speed than the particle moving to the left. Thus, their measured mass would show a difference.

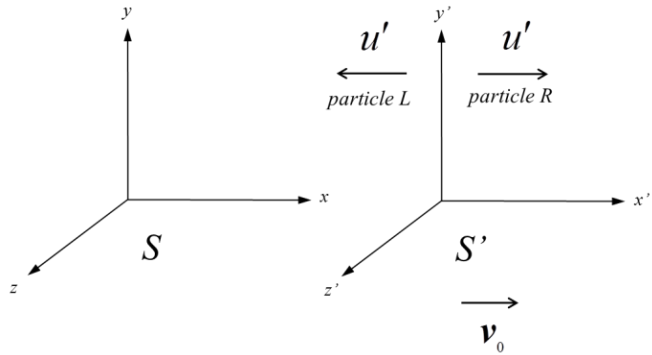

**Fig. 1**. *A schematic diagram showing the relationship between a laboratory frame (S') and a resting frame (S). The frame S' is moving along the x-axis with a velocity $v_0$. The two particles in the S' frame move in opposite directions along the x'-axis with the same speed u'.*

According to the frame **S**, the speed of the particle traveling to the right (*particle R)* is

$$v_R = \frac{u' + v_0}{1 + v_0 u'/c^2}. \quad (2)$$

And the speed of the particle traveling towards the left (*particle L)* is

$$v_L = \frac{-u'+v_0}{1-v_0 u'/c^2}. \tag{3}$$

If both $u'$ and $v_0$ are much smaller than $c$, one can simplify their speeds to

$$\begin{cases} v_R = u'+v_0 \\ v_L = -u'+v_0. \end{cases}$$

Substituting the above relations into Eq. (1) and using Taylor expansion, one can get

$$\frac{\Delta M}{m_0} = \frac{M_R - M_L}{m_0} = \left[1-\left(\frac{v_0+u'}{c}\right)^2\right]^{-1/2} - \left[1-\left(\frac{v_0-u'}{c}\right)^2\right]^{-1/2}$$

$$\approx \frac{1}{2}\left(\frac{v_0+u'}{c}\right)^2 - \frac{1}{2}\left(\frac{v_0-u'}{c}\right)^2 = \frac{2v_0 u'}{c^2}. \tag{4}$$

Thus, by measuring $\Delta M$, one can determine whether the laboratory frame is in motion relative to the resting frame or not.

## 3 Proposed experimental setup

The basic design of our proposed experiment is shown in **Fig. 2**. For simplicity, the charged particles used for this measurement could be electrons. Using an accelerator, the electrons are accelerated to a speed $u'$. The speed of the outgoing electrons can be measured using a TOF (time-of-flight) device. Using a switching magnet, the electrons are directed either to a spectrometer at the right or a spectrometer at the left. The designs of these two mass spectrometers are identical. The masses determined by these two spectrometers ($M_R$ and $M_L$, respectively) are then compared.

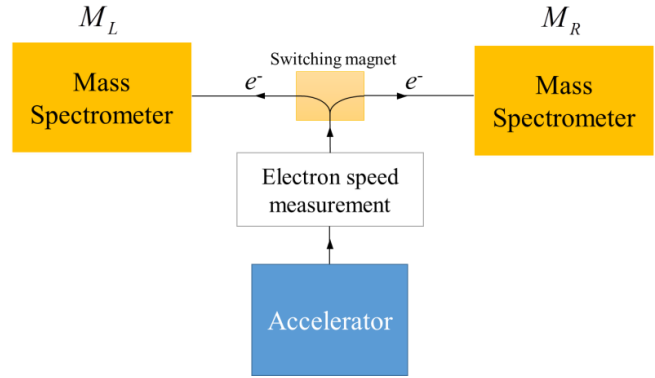

**Fig. 2**. *Conceptual diagram of the experimental setup. Electrons speeded up by an accelerator are analysed by two identical mass spectrometers (located at the left and right). The left-right axis is oriented at the East-West direction.*

This experiment will then be repeated at different time of the day, and in different days of the year. We will examine if any non-zero reading for $\Delta M = M_R - M_L$ can be detected, and whether the measured $\Delta M$ will vary with the orientation of the laboratory frame.

If the 1st postulate of STR is correct, the laboratory frame can be regarded as a stationary frame. Then, $\Delta M$ should always be zero, regardless of the orientation of our experimental setup. If there is a resting frame in our universe, and the moving mass is dependent on its motion relative to this resting frame, we should see a non-zero $\Delta M$; and its value should vary as a function of the orientation of our experimental setup. That means, the measured $\Delta M$ should vary with the hours of the day, and depends on the position of the Earth relative to the Sun.

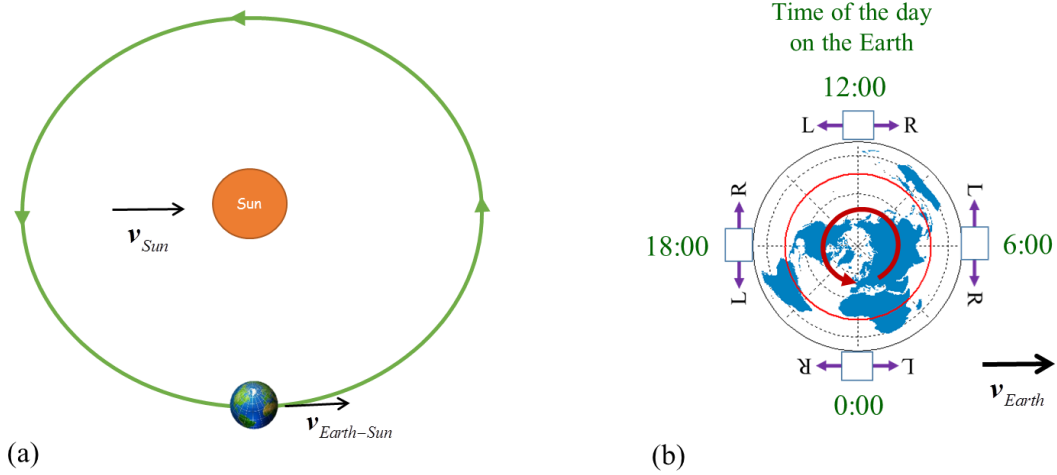

**Fig. 3.** *A simplified diagram showing the basic idea of the experimental design. If there is a resting frame in our universe, it is expected that, for two electrons traveling in opposite directions (right and left), there will be a difference in their moving mass. This mass difference will be seasonal dependent and change with the time of the day. (a) A top view of the movement of the Earth around the Sun. The overall velocity of the Earth ($v_{Earth}$) is a vector sum of the Earth's velocity relative to the Sun ($v_{Earth-Sun}$) and the Sun's velocity relative to the rest of the universe ($v_{Sun}$). Thus, $v_{Earth}$ will change with the season of the year. (b) The R and L arms of the apparatus are pointing to the East-West direction. Because of the movement of the Earth, the orientation of the apparatus is different in relation to $v_{Earth}$ depending on the time of the day. Thus, the electrons moving toward Right and Left will have different velocities relative to the resting frame of our universe. This means that the difference in their moving mass will also change with the hours in a day. (For details, see Table 1.) (Note: For simplicity, we use a 2-D diagram to demonstrate the basic idea here. Two simplifying assumptions have been made: (1) The tilting angle of the Earth's rotation axis is ignored. (2) Here we only consider the Sun's velocity along the plane of Earth's orbit.)*

This point can be seen more clearly from **Fig. 3**, which is a simplified diagram showing the basic idea of our experimental design. **Fig. 3a** is a top view of the orbital motion of the Earth around the Sun. The overall velocity of the Earth ($v_{Earth}$) is a vector sum of the Earth's velocity relative to the Sun ($v_{Earth-Sun}$) and the Sun's velocity relative to the rest of the universe ($v_{Sun}$). Thus, $v_{Earth}$ will change with the season of the year. $v_{Earth}$ will reach a maximum when the Earth reaches a point that $v_{Earth-Sun}$ is roughly parallel to $v_{Sun}$. Since the Earth is continuously rotating, the orientation of the experimental apparatus will change with the time of the day. (See **Fig. 3b**.) Thus, the travelling direction of the electrons (moving toward Right and Left) will have different angles with $v_{Earth}$. For example, at 0:00 hour, $v_L$ is parallel to $v_{Earth}$, while $v_R$ is anti-parallel to $v_{Earth}$. At 12:00 hour, the situation is the opposite: $v_L$ is anti-parallel to $v_{Earth}$, while $v_R$ is parallel to $v_{Earth}$. At 6:00 hour and 18:00 hour, the moving direction of the R and L electrons are both perpendicular to $v_{Earth}$.

Their resulting velocities at different hours are summarized in **Table 1**. Based on the analysis presented in Section 2, one can easily see that, if there is a resting frame in our universe, the moving mass of the R and L electrons should be different, and this difference is expected to vary with the hour of the day.

**Table 1.** *Variation of the moving mass difference depending on the hour of the day*

| **Time** (hour) | $v_R$ | $v_L$ | $\dfrac{M_R - M_L}{m_0}$ |
|---|---|---|---|
| 0:00 | $-u' + v_0$ | $u' + v_0$ | $-\dfrac{2v_0 u'}{c^2}$ |
| 6:00 | $\sqrt{(u')^2 + v_0^2}$ | $\sqrt{(u')^2 + v_0^2}$ | 0 |
| 12:00 | $u' + v_0$ | $-u' + v_0$ | $\dfrac{2v_0 u'}{c^2}$ |
| 18:00 | $\sqrt{(u')^2 + v_0^2}$ | $\sqrt{(u')^2 + v_0^2}$ | 0 |

*Note: Because the motion from the Earth's rotation is much slower than the velocity of the Earth ($v_{Earth}$), the speed of the laboratory frame ($v_0$) is essentially the same as $v_{Earth}$.*

Now, let us consider whether it is feasible to measure $\Delta M / m_0$ using the existing technology. This depends on how large $v_0$ could be. Since our laboratory is located on the planet Earth, we can estimate the value of $v_0$ from the relative motions of the planet and our solar system. The speeds of those motions are known at present (see **Table 2**).

**Table 2.** *The speeds of the relative motions of the Earth, solar system and Milky Way*

| **Type of movement** | **Speed (m/s)** | **Reference** |
|---|---|---|
| Motion from the Earth's rotation | $v_{Rotation} = 465.1$ | [10] |
| Motion relating to Earth's orbiting around the Sun | $v_{Earth-Sun} = 2.98 \times 10^4$ | [10, 11] |
| Motion due to the Solar system orbiting around the Galaxy | $v_{Sun-Gal} \sim 2.2 \times 10^5$ | [12] |
| Solar system moving in reference to CMB | $v_{Sun-CMB} \sim 3.7 \times 10^5$ | [13, 14] |

From Table 2, one can see that the major contributions to the speed of the laboratory frame are due to $v_{Sun-Gal}$ and $v_{Sun-CMB}$. By comparison, the contribution by the motion from the Earth's rotation is negligible (less than 0.2%). The overall velocity of the Sun's movement ($v_{Sun}$) is a vector sum of $v_{Sun-Gal}$ and $v_{Sun-CMB}$. Since the directions of these movements are not parallel, one may roughly estimate that $v_{Sun}$ could be in the order of $3 \times 10^5$ m/s. Since the speed of the Earth ($v_{Earth}$) is dominated by $v_{Sun}$, and the speed of the laboratory frame (on the Earth surface) is dominated by $v_{Earth}$, this means that the value of $v_0$ in **Fig. 1** is also in the order of $3 \times 10^5$ m/s, i.e., $v_0/c \approx 10^{-3}$. In our proposed experiment, suppose we can accelerate the electron to 1/10 of the speed of light, i.e., $u'/c = 0.1$, then

$$\frac{\Delta M}{m_0} = \frac{2v_0 u'}{c^2} \approx 2 \times 10^{-4}. \tag{5}$$

This means that, during the best days of the year, the observable value of $\Delta M / m_0$ could oscillate from $+2 \times 10^{-4}$ to $-2 \times 10^{-4}$ within 24 hours. Such a data range is not too small and should be measurable using the existing experimental techniques.

## 4 Design of the directional mass spectrometer

As to the design of the mass spectrometer, it must satisfy two requirements: (1) It must be precise enough to determine the mass with an accuracy better than $10^{-5}$. (2) During the measurement, the orientation of the particle cannot deviate too much from the direction of the incident beam. Otherwise, the moving mass may change due to the change of speed in the resting frame. At present, the most sensitive measurements of the electron mass-charge ratio are conducted using electrons stored in a Penning trap [15, 16]. It could achieve an accuracy of about $10^{-8}$, which is more than we need. Unfortunately, such an experimental design cannot be used in our proposed experiment. Since trapped electrons must undergo constant circular motion, their movement cannot be maintained in a fixed direction with the resting frame. Thus, it will not fit our requirement.

Thus, we need to design a mass spectrometer which does not depend on measuring the resonance frequency of the trapped electron. **Fig. 4** shows the conceptual diagram of a workable mass spectrometer designed for our proposed experiment. The electron beam passing through a magnetic field (**B**) is slightly deflected at an angle $\theta$. This deflection angle $\theta$ depends on three parameters: (1) the strength of the magnetic field, (2) the speed of the electron, and (3) the mass of the electron. Since **B** is maintained constant in this experiment and the speed of the electron entering the mass spectrometer is also fixed (in the laboratory frame), we can determine the mass of the electron by measuring the deflection angle $\theta$.

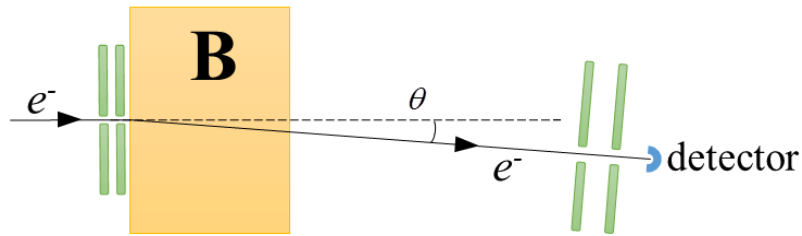

**Fig. 4**. *A conceptual diagram showing the basic design of the mass spectrometer used in this experiment. The incoming electron is deflected by a magnetic field (**B**) and exits at an angle $\theta$. The electron is detected by a highly sensitive image sensor that can accurately measure the position of the deflected electron.*

In this experiment, a highly sensitive electron image detector is placed behind the detection collimator. One can use a CMOS sensor as such a detector [17]. Using this imaging device, the angle $\theta$ can be determined with high accuracy. For example, using state-of-the-art microfabrication technique at present, the pixel of the sensor can be made small enough to achieve a resolution in the order of 1 $\mu m$ [17, 18]. If the distance between the entrance collimator and the electron detector is set at 5 $m$, the deflection angle $\theta$ can be measured with an accuracy of about $0.2 \times 10^{-6}$ radian. In the experiment, one may adjust the magnetic field so that deflection angle for the electron is in the range of 0.2 radian. (Note: In order to satisfy our requirement (2), the angle $\theta$ must be kept small in the experiment.) Since the deflection angle is inversely proportional to the mass of the particle, the relative accuracy of the mass measurement is the same as the accuracy of the deflection angle measurement, i.e., $\delta M / M \approx \delta\theta / \theta \approx 10^{-6}$ (here $\delta\theta$ and $\delta M$ are the measurement uncertainty for $\theta$ and $M$, respectively). Such accuracy is more than enough to detect the expected mass difference $\Delta M$ if there is a resting frame.

Without saying, the electron beam in this experiment is traveling in high vacuum. The electron source may be cooled to reduce the spread of momentum. One may ask that: Can the magnetic field and the electron speed be maintained to the same accuracy as the measurement of the deflection angle $\theta$ (i.e., better than $10^{-6}$) during the experiment? This should not be a problem. First, using a NMR device, one can regulate the magnetic field to an accuracy better than $10^{-8}$ [19]. Second, since electrons entering the left and right spectrometers are produced from the same electron accelerator, their speeds should be identical. Third, even if the magnets used in the left and right spectrometers are not perfectly matched, it can only produce a systematic error that would shift the baseline of the $\Delta M / m_0$ measurement. It will not affect the cyclic behavior of the data measured over one day.

## 5 Discussions

This proposed experiment is not very complicated. It can be conducted using existing experimental techniques. For simplicity, one can use electron as the charged particle in the experiment. Of course, the same experiment can be done with protons, which may have the advantage of higher precision, since the proton has a larger rest mass and so it may be easier to detect the $\Delta M$.

One may see that the design of this experiment is partially inspired by the optical interferometer experiments previously conducted by Michelson and Morley [1]. Their experiment demonstrated that for particles with no rest mass, i.e., photons, their behaviour obey the 1st postulate of STR. Our experiment is designed to further test whether particles with nonzero rest mass can also satisfy the same principle.

One may question that, since our experimental design is based on assuming the validity of Eq. (1), which is usually attributed to STR, does our starting point already assume that the 1st postulate of STR is correct? This is not the case. Strictly speaking, the validity of Eq. (1) is more dependent on experimental agreement than its theoretical derivation [9, 20]. The derivation of Eq. (1) actually does not require the 1st postulate of STR [20]. Furthermore, the derivation of Eq. (1) is not unique. It had been shown that one can derive Eq. (1) from a Matter Wave model in which the particle is regarded as an excitation wave of the vacuum [21, 22]. Thus, there is no logical contradiction in using Eq. (1) to test if there is a resting frame in the universe or not.

The physical properties of the vacuum play a major role in many models of cosmology today [23, 24]. It is very important to know whether the vacuum can serve as a resting frame. In this paper, we propose a simple method to investigate this outstanding problem. The idea proposed here is straight forward. Its experimental requirements are within the current technological capabilities. In fact, the major equipment mentioned here are readily available in many low-energy nuclear physics laboratories. Thus, the idea presented in this work can easily be developed into a well-designed experimental project. This could be a very efficient way to test a fundamental problem in physics.

I may add that it has been a long history in physics to search for experimental evidence to test the limit of the first postulate of STR. Recently, such search is further motivated by a new development, i.e., many current models of quantum theory for gravity have a feature of violation of Lorentz invariance. Thus, there have been attempts to conduct new experiments to test the limits of STR. For example, Muller et al. had proposed to study possible violations of Lorentz invariance based on electrons in a crystal [25]. This work represents the most recent attempt, which is based on a very clear physical foundation.

## 6 Conclusion

This work propose an experiment to test two basic scenarios: *(1) There is no resting frame in our universe*; *(2) There is a resting frame in our universe*. In this proposed experiment, *Scenario (1)* would predict that the $\Delta M / m_0$ measurement should always give a null result, which is regardless of the time of the day or the day of the season. *Scenario (2)*, on the other hand, would predict that the $\Delta M / m_0$ measurement will give a non-zero result, the value of which will be strongly dependent on the time of the day (and the day of the season). The amplitude of its variation is expected to be in the order of $\frac{\Delta M}{m_0} = \frac{2v_0 u'}{c^2} \approx 2 \times 10^{-4}$. In this proposed experiment, we will compare the measurement results directly with the theoretical predictions of those two scenarios. Our analysis shows that, with the existing technological limits, the proposed measurement should have sufficient accuracy to differentiate the two scenarios.

**References:**

[1]   A. A. Michelson, E. W. Morley, On the relative motion of the Earth and the luminiferous ether. *Am. J. Sci.*, **34**, 333-345 (1887).
[2]   A. Einstein, *The Principle of Relativity: A Collection of Original Memoirs on the Special and General Theory of Relativity* (Dover Publications, New York, 1952).
[3]   A. Einstein, *Relativity: The Special and General Theory* transl. R. W. Lawson, (Pi Press, New York, 2005), p 19.
[4]   A. H. Guth, D. I. Kaiser, Inflationary cosmology: Exploring the universe from the smallest to the largest scales. *Science*, **307**, 884-890 (2005).
[5]   C. L. Bennett *et.al.*, Nine-Year Wilkinson Microwave Anisotropy Probe (WMAP) observations: Final maps and results. *Astrophys. J. Supp. Series*, **208**, 20B (2013).
[6]   Plank Collaboration, R. Adam *et.al.*, Planck 2015 results. *Astronomy & Astrophys.*, **594**, A1 (2016).
[7]   A. Messiah, *Quantum Mechanics* (Wiley, New York, 1965) p 439.
[8]   L. H. Ryder, *Quantum Field Theory*, (Cambridge Univ. Press, New York, 1996) p 284.

[9] P. S. Faragó, L. Jánossy, Review of the experimental evidence for the law of variation of the electron mass with velocity. *Nuovo Cim,* **5**, 1411 (2008).
[10] A. N. Cox, Ed., *Allen's Astrophysical Quantities*, (AIP Press; Springer, New York, 2000) pp. 244-245.
[11] NASA Earth Fact Sheet. (19 May 2016). Available: http://nssdc.gsfc.nasa.gov/planetary/factsheet/earthfact.html.
[12] M. A. Garlick, *The Story of the Solar System*, (Cambridge Univ. Press, New York, 2002) p 46.
[13] G. Hinshaw *et al.*, Five-year Wilkinson Microwave Anisotropy Probe observations: Data processing, sky maps, and basic results. *Astrophys. J. Supp. Series*, **180**, 225-245 (2009).
[14] A. Kogut *et al.*, Dipole anisotropy in the COBE differential microwave radiometers first-year sky maps. *Astrophys. J.*, **419** (1993).
[15] K. Blaum, Y. N. Novikov and G. Werth, Penning traps as a versatile tool for precise experiments in fundamental physics. *Contemp. Phys.,* **51**, 149-175 (2010).
[16] S. Sturm, F. Köhler, J. Zatorski, A. Wagner, Z. Harman, G. Werth, W. Quint, C. H. Keitel and K. Blaum, High-precision measurement of the atomic mass of the electron. *Nature,* **506**, 467-470 (2014).
[17] M. Artuso *et al*, Sensor compendium. *arXiv:1310.5158 [Physics. Ins-Det].*
[18] M. Battaglia *et al*, Characterisation of a CMOS active pixel sensor for use in the TEAM microscope. *Nucl. Instrum. Methods Phys. Res. Section A-Accelerators Spectrometers Detectors Associated Equipment,* **622**, 669-677 (2010).
[19] C. P. Slichter, *Principles of Magnetic Resonance* (Springer-Verlag, Berlin; Hong Kong, 1990).
[20] A. P. French, *Special Relativity*, (Norton, 1968) pp. 20-28.
[21] D. C. Chang, Why energy and mass can be converted between each other? A new perspective based on a matter wave model. *J. Mod. Phys.*, **7**, 395-403, (2016).
[22] D. C. Chang, What is the physical meaning of mass in view of wave-particle duality? A proposed model. *arXiv: 0404044v2 [physics.gen-ph].*(2016).
[23] S. Kachru, R. Kallosh, A. Linde and S. P. Trivedi, de Sitter, Vacua in string theory. *Phys. Rev. D*, **68**, 046005, (2003).
[24] A. H. Guth, D. I. Kaiser and Y. Nomura, Inflationary paradigm after Planck 2013. *Phys. Lett. B*, **733**, 112-119, (2014).
[25] H. Mueller et al., Optical cavity tests of Lorentz invariance for the electron. *Phys. Rev. D,* **68**, 116006, (2003)

**Acknowledgements:** I thank Ms. Lan Fu for her assistance. This work was partially supported by the Research Grant Council of Hong Kong (RGC 660207) and the Macro-Science Program, Hong Kong University of Science and Technology (DCC 00/01.SC01).